\documentclass{article}

\usepackage{arxiv}
\usepackage[utf8]{inputenc} 
\usepackage[T1]{fontenc}    
\usepackage{hyperref}       
\usepackage{url}            
\usepackage{booktabs}       
\usepackage{amsfonts}       
\usepackage{nicefrac}       
\usepackage{microtype}      
\usepackage{graphicx}
\usepackage{doi}
\usepackage{authblk}
\usepackage{amsmath}

\title{Covariant field lines: a geometrical approach to electrodynamics}

\author[1, $\dag$]{Yaron Hadad}
\author[1]{Ido Kaminer}
\author[2,3]{Avshalom Elitzur}
\author[4]{Eliahu Cohen}

\affil[1]{Technion, Israel Institute of Technology, Haifa, Israel}
\affil[2]{Iyar, The Israeli Institute for Advanced Research, Zichron Ya'akov, Israel}
\affil[3]{Chapman University, Orange, California, USA}
\affil[4]{Bar Ilan University, Ramat Gan, Israel}
\affil[$\dag$]{e-mail: yaron.hadad@gmail.com}

\hypersetup{
pdftitle={Covariant Field Lines: A Geometrical Approach to Electrodynamics},
pdfsubject={math-ph, physics.class-ph},
pdfauthor={Yaron Hadad, Ido Kaminer, Avshalom Elitzur, Eliahu Cohen},
pdfkeywords={electrodynamics, field lines, Lorentz force, electrodynamic equations of motion, self force, radiation reaction},
}

\begin{document}
\maketitle

\begin{abstract}

This paper revisits the geometric foundations of electromagnetic theory, by studying Faraday's concept of field lines. We introduce "covariant electromagnetic field lines," a novel construct that extends traditional field line concepts to a covariant framework. Our work includes the derivation of a closed-form formula for the field line curvature in proximity to a moving electric charge, showcasing the curvature is always non-singular, including nearby a point charge. Our geometric framework leads to a geometric derivation of the Lorentz force equation and its first-order corrections, circumventing the challenges of self-force singularities and providing insights into the problem of radiation-reaction. This study not only provides a fresh geometric perspective on electromagnetic field lines but also opens avenues for future research in fields like quantum electrodynamics, gravitational field theory, and beyond.

\end{abstract}


\section{Introduction}

The notion of fields is ubiquitous in modern physics, both classical and quantum. It was Michael Faraday who first introduced this idea in his work on classical electromagnetism. Faraday studied fields by exploring the concept of \emph{field lines} (also known as \emph{lines of forces}), which he applied to electric, magnetic and gravitational fields. For Faraday, as well as for Maxwell later, field lines were real physical entities; electric and magnetic forces were the manifestation of the stress that field lines induce \cite{bib:Faraday}.

With the introduction of special relativity, electric and magnetic fields were unified as a single field \cite{bib:Einstein}, making the distinction between electric and magnetic field obsolete. This is one of the reasons that today field lines are mostly used as a pedagogical tool, although they are also helpful for describing a variety of complex phenomena \cite{bib:Wald74,bib:King75,bib:Bicak80,bib:Kom07,bib:Bini08, bib:Kom04,bib:Bini07, bib:Lichnerowicz, bib:Anile}. 

As long as fields can be separated to electric or magnetic, we may picture them using their field lines. Even then, they are usually applied only to electrostatic or magnetostatic physical systems. It would be valuable to extend the theory of field lines to a fully relativistic framework.

\begin{figure} [ht]
\begin{center}
\includegraphics[scale=.5]{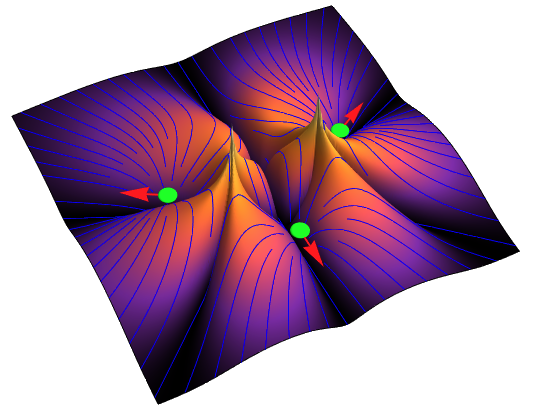}
\caption{\emph{Charges accelerate in the direction of straight field lines}: Electromagnetic field lines are represented by curves (in blue), whose curvature is represented by the color and elevation of the surface (bright colors correspond to higher curvature/elevation). In this figure, three electrons (green) repel one another in the direction of the (red) arrows. The curvature is usually non-zero nearby each charge, yet the field lines pointing in the direction of the charge's acceleration are straight (zero curvature). Charges always accelerate in the direction of straight electromagnetic field lines, in analogy with the motion along a geodesic in a gravitational field in general relativity. The two peaks in the figure are singular points, which only occur where the total electromagnetic field is zero (since no field lines pass through these points).
\label{fig:FieldLinesCurvature}}
\end{center}
\end{figure}

In this work, we propose a new formalism of covariant electromagnetic field lines treating electric and magnetic fields on equal footing. The resulting covariant field lines naturally generalize both electric and magnetic field lines. The covariant field lines introduced are curves in Minkowski spacetime, and their geometrical properties reveal meaningful physical consequences. In particular, we calculate the curvature of the field lines and use it to derive the charge's equation of motion from first principles. Interestingly, the field lines bend due to external electromagnetic fields or charge acceleration, yet charges always travel along straight field lines (Fig. \ref{fig:FieldLinesCurvature}). We also show that the field line curvature is well-defined and non-singular wherever the electromagnetic field does not vanish, even in the vicinity of point charges. Thus, the covariant field lines not only help visualize the electromagnetic force, but also offer new tools to better understand and eventually solve fundamental electromagnetic problems such as self-force and radiation reaction arising in the presence of strong fields as created by high-intensity lasers. 

Geometric approaches to electromagnetism were studied extensively in the literature. Previous attempts on geometrization of electromagnetism include Kaluza-Klein theory \cite{bib:Kaluza, bib:Klein}, Rainich-Misner-Wheeler theory \cite{bib:Rainich, bib:MisnerWheeler}, Yang-Mills theory \cite{bib:YangMills}, and Gauge theory \cite{bib:Atiyah, bib:Straumann}. The covariant field lines approach given here is fundamentally different from previous attempts, as we do not geometrize space or spacetime. Instead, it is the field lines themselves that are treated as geometrical objects and the field lines curvature is the focus of this work.

The paper is organized as follows. In sections 2 and 3 we outline our construction of covariant electromagnetic field lines and their curvature. We then discuss in sections 4 the consequences of this formalism, including two fundamental consequences - the tight relation between acceleration and curvature, and the absence of curvature singularities at the vicinity of charges, respectively. We conclude in section 5, and also discuss how the field line formalism can derive the Lorentz Force equation and potentially shed light on radiation reaction equations.

\section{Covariant field lines} \label{sec:CovariantFieldLines}

The electric field lines $\vec{x}_e (\lambda)$ and magnetic field lines $\vec{x}_m (\lambda)$ for an arbitrary parametrization $\lambda$ are given by the electric and magnetic field line equations
\begin{equation} \label{eq:3DFieldLines}
\frac{d \vec{x}_e}{d\lambda} = \vec{E}(\vec{x}_e(\lambda)) \hspace{1cm} \frac{d \vec{x}_m}{d\lambda} = \vec{B}(\vec{x}_m(\lambda)).
\end{equation}
It has been argued \cite{bib:Newcomb,bib:VanEnk} that electric and magnetic field lines are inherently non-relativistic for two reasons. First, since electric and magnetic fields transform into each other under Lorentz transformations one cannot expect to find a covariant notion of a pure electric or a pure magnetic field line. Second, points on electric (or magnetic) field line all have the same time coordinate and are therefore inherently non-local.

In Minkowski spacetime the electromagnetic field is represented by the electromagnetic tensor $F^{\mu\nu}$, a covariant tensor that treats the electric and magnetic fields as a single physical entity. One would hope that the first issue raised above could be addressed by defining the electromagnetic field lines using the tensor $F^{\mu\nu}$. This way, instead of splitting the electric and magnetic field lines we consider them as a single entity resulting from a covariant electromagnetic field. The second aforementioned issue can be overcome by refraining from the notion of absolute time in the definition of covariant field lines. Instead, we define the electromagnetic field lines relatively to the charge that is measuring them (i.e. the test charge).

\begin{figure}
\begin{center}
\includegraphics[scale=.5]{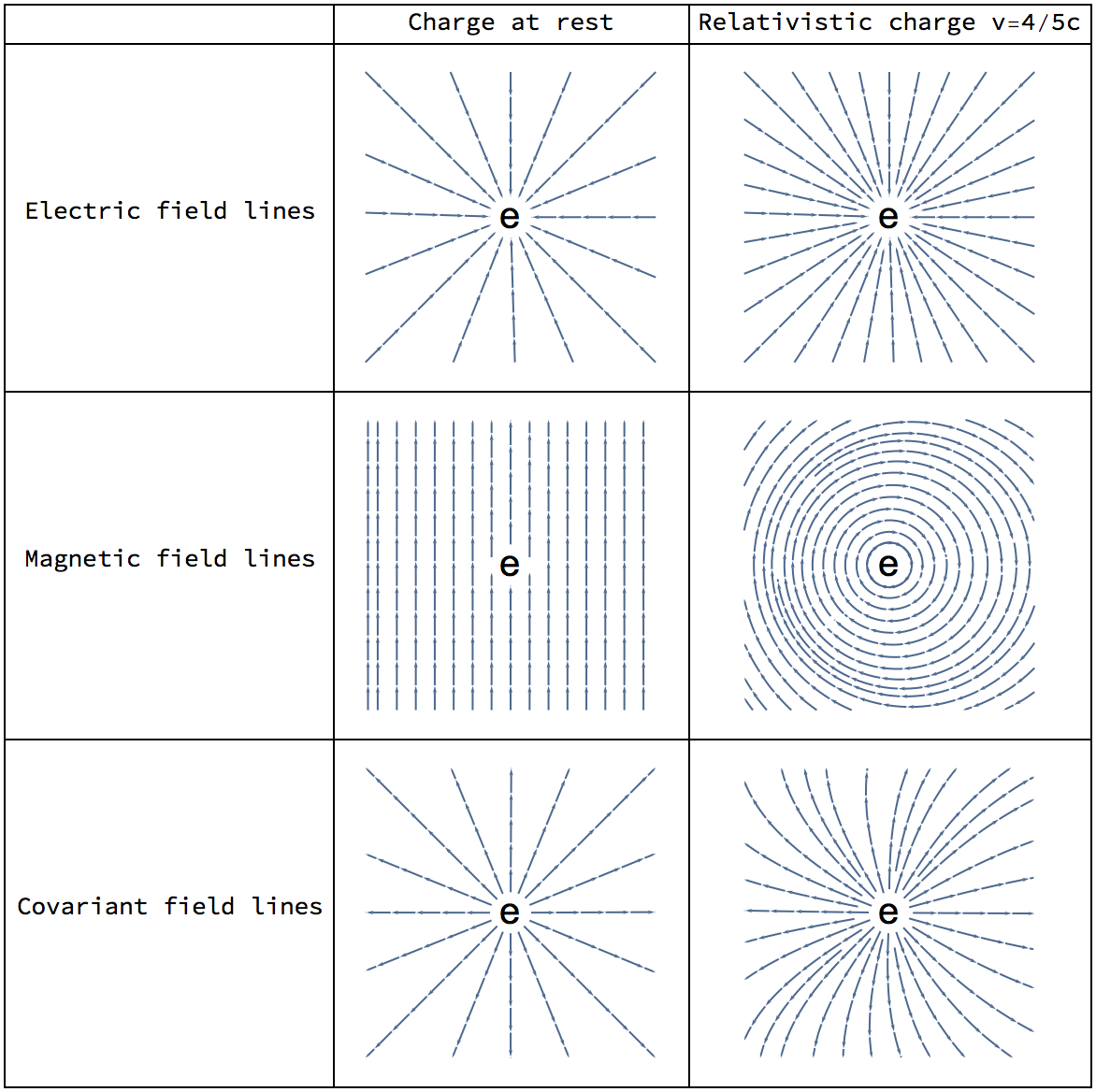}
\caption{\emph{Illustrations of electric, magnetic, and covariant field lines}: The total electric, magnetic and covariant electromagnetic field lines are presented for an electron at rest (first column) and an electron at a relativistic velocity $v=4/5c$ out of the page (second column). The external field is a constant magnetic field upwards (no electric field), as can be seen in the second figure from the top (resting charge). For a charge at rest, the electric and covariant field lines coincide. Once the charge is moving, the electric field lines are denser in the covariant direction (well known), while the covariant field lines are both denser and also curve due to the magnetic force. The relativistic electron is expected to accelerate in the horizontal direction as the covariant field lines are flat in that direction.
\label{fig:EBCPanel}}
\end{center}
\end{figure}

We define \emph{the covariant electromagnetic field line equation}:
\begin{equation} \label{eq:CovariantFieldLines}
\frac{d x ^\mu}{ds} = F^{\mu\gamma} (x^\nu) u_\gamma (\tau_\text{ret}),
\end{equation}
where $F^{\mu\nu}$ is the electromagnetic field tensor, $u^\gamma(\tau_\text{ret})$ is the four-velocity of the test charge that is evaluated at the retarded time $\tau_\text{ret}$. We use a spacetime metric with signature $(-+++)$. It may appear that this is nothing but the Lorentz force equation, but this is not the case. The covariant field line Eq. (\ref{eq:CovariantFieldLines}) is a first-order differential equation (and not second-order), which unlike the Lorentz force equation is defined using the retarded time $\tau_\text{ret}$. In fact, it is shown in the Appendix that Eq. (\ref{eq:CovariantFieldLines}) is the \emph{unique field lines equation} that is covariant, linear in the electromagnetic field, and generalizes the electric and magnetic field lines Eqs (\ref{eq:3DFieldLines}) to the relativistic case. Moreover, one can show that this defines the covariant field lines throughout spacetime and that this equation is well-posed. Covariant electromagnetic field lines indeed generalize electric and magnetic field lines, but also have one main differentiating factor. They depend on the charge that is measuring them (i.e. probe or test charge), as is expected from a relativistic theory. In contrast, both electric and magnetic field lines are velocity- and observer-independent (Fig. \ref{fig:EBCPanel}).

\section{The curvature of a field line}\label{sec:FieldLinesCurvature}

\begin{figure}
\begin{center}
\includegraphics[scale=.25]{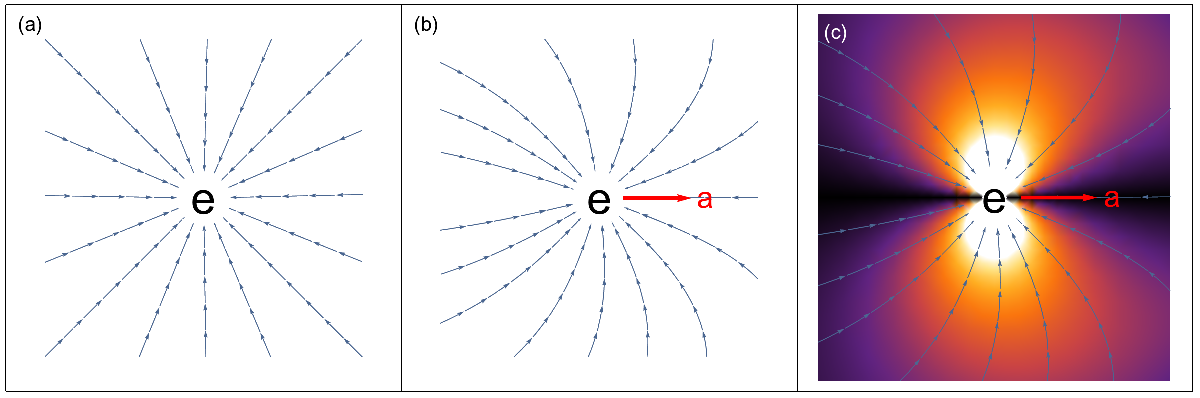}
\caption{\emph{Correspondence between charge acceleration and field line curvature} (a) The field lines of a free charge at rest. (b) As the charge accelerates (red arrow), the field lines bend. (c) The field line curvature determines the acceleration of the charge. The charge always accelerates in the direction of straight field lines (black color represent zero curvature). Straight field lines are not unique, and there can be multiple straight field lines around the charge.
\label{fig:FieldLinesOfAcceleratingCharge}}
\end{center}
\end{figure}

The authors have previously studied the geometry of \emph{electric} field lines and showed it is related to the dynamics of non-relativistic charges \cite{bib:Hadad1,bib:Hadad2}. As a charge accelerates the field lines around it bend. Despite this, the field line in the direction of acceleration remain straight (Fig. \ref{fig:FieldLinesOfAcceleratingCharge}). Conversely, one can prove that if the electric field lines bend due to an external field, a charge will accelerate in the direction of straight field lines. This gives rise to an appealing relationship:
\begin{equation*}
\text{field line curvature} \leftrightarrow \text{charge acceleration}.
\end{equation*}
The derivation of this relationship in electrostatics and magnetostatics is provided in the Appendix \ref{sec:ElectrostaticFieldLines}. As we show below, this relationship holds for a general electromagnetic field.

In differential geometry, a curvature vector of a curve is defined as the curve's second derivative with respect to its unit-length parametrization $s$. In the case of the covariant field line equation (\ref{eq:CovariantFieldLines}), its four-curvature vector may be written explicitly as the difference:
\begin{equation} \label{eq:VectorCurvature}
\kappa^\mu = \frac{1}{G^2} G^\nu \frac{\partial G^\mu}{\partial x^\nu} - \frac{1}{G^4} (G_\gamma G^\nu \frac{\partial G^\gamma}{\partial x^\nu}) G^\mu,
\end{equation}
where $G^\mu (x^\nu) = F^{\mu\gamma} (x^\nu) u_\gamma (\tau_\text{ret})$ as was defined in the Appendix and we used the chain-rule $\frac{d}{ds} = \frac{d x^\gamma}{d s} \frac{\partial}{\partial x^\gamma}$.

The reader can easily verify that for a charge at rest, the spatial components the curvature four-vector in Eq. (\ref{eq:VectorCurvature}), and the scalar curvature $\kappa = \sqrt{\kappa^\mu \kappa_\mu}$ give the same result as the electrostatic limit provided in the Appendix. Therefore, the covariant field lines and their curvature as defined here naturally extend their non-relativistic form in Eq. (\ref{eq:3DFieldLines}) used in \cite{bib:Hadad1, bib:Hadad2}.


\begin{figure} [h]
\begin{center}
\includegraphics[scale=.75]{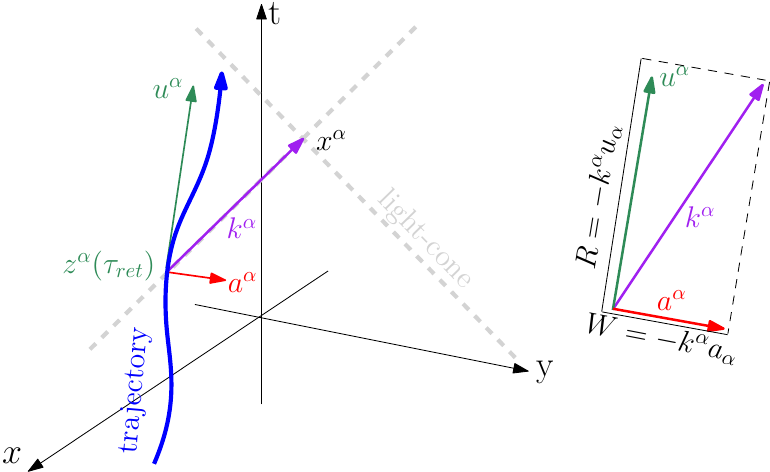}
\caption{\emph{Useful notations}: The world line of the charge $q$ in spacetime, where the null-vector $k^\mu$ connects the charge's retarded four-position $z^\mu (\tau_{\text{ret}})$ to an arbitrary event $x^\mu$. The retarded four-velocity of the charge $u^\mu$ is the tangent to the world line and is perpendicular to the retarded four-acceleration $a^\mu$.
\label{fig:LienardWiechert}}
\end{center}
\end{figure}

We now proceed to study the curvature of the field lines. Consider a charge $q$ with mass $m$ traveling along a world line $z^\mu(\tau)$ with proper time $\tau$ (Fig. \ref{fig:LienardWiechert}). The total electromagnetic field can be written as a superposition $F^{\mu\nu} = F^{\mu\nu} _\text{ext} + F^{\mu\nu} _\text{self}$ of the external field $F^{\mu\nu} _\text{ext}$ and the self field $F^{\mu\nu} _\text{self}$ produced by the charge itself on its forward light-cone. A point $x^\mu$ is on the light-cone of the charge if there exists $\tau_{\text{ret}}$ for which $k^\mu k_\mu = 0$ for $k^\mu = x^\mu - z^\mu (\tau _\text{ret})$. The self field produced by the charge is given by the retarded Li\'enard-Wiechert equation \cite{bib:Rohrlich}
\begin{equation} \label{eq:Lienard-Wiechert}
F^{\mu\nu} _{\text{self}} = \frac{q}{R^2} \left(U^\mu k^\nu - U^\nu k^\mu\right),
\end{equation}
where $R=-k^\mu u_\mu$ and $W=-k^\mu a_\mu$ are the retarded distance and retarded velocity respectively. $U^\mu= B u^\mu + a^\mu$ is the Synge vector \cite{bib:Synge} and $B=\frac{1-W}{R}$ is the Pleba\~nski invariant \cite{bib:Plebanski}. The four-velocity and four-acceleration used in Eq. (\ref{eq:Lienard-Wiechert}) are the retarded four-velocity and four-acceleration $u^\mu = u^\mu (\tau_\text{ret})$ and $a^\mu = a^\mu (\tau_\text{ret})$ respectively.

We are interested in finding the influence of the curvature on a charge, and therefore look for the curvature in the vicinity of the charge. To do so, recall from Eq. (\ref{eq:Lienard-Wiechert}) that the self field at each event $x^\mu$ is determined by the four-position of the charge at the retarded four-position $z^\mu (\tau_\text{ret})$. This means that one needs to study the limit $x^\mu \rightarrow z^\mu$ along the light-cone. To do, let $k^\mu = \varepsilon \hat{k}^\mu$ where $\hat{k}^\mu$ is of order unity and $\varepsilon$ is a small parameter. In practice, this is equivalent to setting $\varepsilon=k^0$ and taking the limit $\varepsilon \rightarrow 0$.

The curvature four-vector in Eq. (\ref{eq:VectorCurvature}) can be expanded in $\varepsilon$, and gives up to the first order
\begin{equation} \label{eq:Curvature}
\begin{array}{l}
\kappa^\mu = \left[-\frac{\hat{W}}{\hat{R}^2} \hat{P}^\mu - a^\mu + \frac{\hat{W}}{\hat{R}} u^\mu \right] +\\
\hspace{.8cm} \varepsilon \left[\left(\frac{2\hat{W}^2}{\hat{R}^2} - 2a^2-\frac{3}{q\hat{R}} \hat{k}_\nu F_{\text{ext} }^{\nu\gamma} u_\gamma + \frac{\hat{A}}{\hat{R}}\right)\hat{P}^\mu - \left(\frac{\hat{W}^2}{\hat{R}} u^\mu + \hat{W} a^\mu + \hat{R} \dot{a}^\mu + \frac{3 \hat{R}}{q} F_{\text{ext}} ^{\mu\nu}u_\nu  \right)\right]
\end{array}
\end{equation}
where parameters denoted by a hat are of order unity, namely $k^\mu = \varepsilon \hat{k}^\mu$, $W=\varepsilon \hat{W}$, $R=\varepsilon \hat{R}$, $\hat{P}^\mu = \hat{k}^\mu - \hat{R} u^\mu$, and $\hat{A} = -\hat{k} \cdot \dot{a}$. Additional details on the derivation are provided in the Appendix.

Eq. (\ref{eq:Curvature}) gives a direct way of calculating the curvature of the field lines nearby the charge $q$ in a straightforward manner. We remind the reader that all the quantities on the right-hand side of Eq. (\ref{eq:Curvature}) are defined using the retarded time, and in particular the four-velocity and four-acceleration used in Eq. (\ref{eq:Curvature}) are the retarded four-velocity $u^\mu = u^\mu (\tau_\text{ret})$ and retarded four-acceleration $a^\mu = a^\mu (\tau_\text{ret})$. This dependence on $\tau_\text{ret}$ means that the right-hand side is defined wherever its denominator $\hat{R}$ is non-zero, and in particular one may consider $\kappa^\mu$ as a vector field in spacetime. We will show in the next section that this field is in fact defined throughout spacetime and does not exhibit any singularities.

\section{Consequences}
\subsection{The correspondence between curvature and acceleration}\label{subsec:CurvatureIsAcceleration}

Geometrically, the four-acceleration of a physical object is the curvature vector of its world line \cite{bib:Pauli}. This is an equivalence between the kinematics of the object and the geometry of its world line, which links acceleration (of the object) and curvature (of its world line). An immediate consequence of Eq. (\ref{eq:Curvature}) is that an object's acceleration can also be said to be equivalent to its electromagnetic field line curvature (if the object is charged). To see this, consider the four-curvature of the field lines in Eq. (\ref{eq:Curvature}) to the leading order in $\varepsilon$:
\begin{equation} \label{eq:VectorCurvatureFreeCharge}
\kappa ^\mu = -a^\mu -\frac{\hat{W}}{\hat{R}^2} \hat{k}^\mu + \frac{2\hat{W}}{\hat{R}} u^\mu + O(\varepsilon),
\end{equation}
with squared length $\kappa^2 = a^\mu a_\mu - \frac{2\hat{W}^2}{\hat{R}^2}$, both of which are independent of the external force. In the rest frame of the charge $a^\mu = (0,\vec{a})$ and the above equation reduces to $\kappa^2 = -\vec{a}^2 \cos (2\theta)$, where $\theta$ is the angle between the spatial acceleration $\vec{a}$ and the spatial part of the null vector $\hat{k}$. Therefore, in the leading order, the scalar field line curvature is maximal and of the same magnitude of the spatial acceleration precisely when $\theta=\pi/2$ ($\hat{W}=0$). This is the same angle $\theta$ for which there is a maximal angular emission of radiation \cite{bib:Jackson}. 

The condition $\hat{W}=-\hat{k} \cdot a = 0$ defines a two-dimensional surface in spacetime that intersects the four-position of the charge. On this surface Eq. (\ref{eq:VectorCurvatureFreeCharge}) gives $\kappa ^\mu = -a^\mu$. This implies that the four-acceleration of the charge is precisely (minus) the curvature of the total electromagnetic field lines in its reference frame. This also implies that we may interchangeably translate a property related to the kinematics of the charge (acceleration) and the geometry of the field lines (curvature). We therefore get a sequence of three equivalences in classical electrodynamics, relating four-acceleration, curvature of world lines and curvature of electromagnetic field lines:
\begin{equation}
\text{four-acceleration } a^\mu = \text{curvature of world line} = - \text{curvature of field lines}.
\end{equation}

\subsection{No singularities nearby charges} \label{subsec:NoSingularities}

As is well known as ``the self-field problem'', the electromagnetic field tensor defined in Eq. (\ref{eq:Lienard-Wiechert}) is singular at the position of the charge and behaves like $\frac{1}{R^2}$ in the limit $R\rightarrow 0$ \cite{bib:Jackson, bib:Griffiths}. It may appear that Eq. (\ref{eq:Curvature}) or Eq. (\ref{eq:VectorCurvatureFreeCharge}) are singular due to the denominators containing the normalized retarded position $\hat{R}$, but this is not the case and in fact it is a removable singularity.

To see this, we explicitly write the normalized null-vector $\hat{k}^\mu = (1,\hat{k})$ and the four-velocity $u^\mu=\gamma(1,\vec{v})$, where $\hat{k}$ is a unit spatial vector, $\vec{v}$ is the spatial velocity of the charge in units where the speed of light is unity $c=1$, and $\gamma=\frac{1}{\sqrt{1-\vec{v}^2}}$ is the relativistic Lorentz factor. This shows that the normalized retarded distance is never zero $\hat{R} = \gamma (1 - \hat{k}\cdot \vec{v})\neq 0$, as the charge is not massless and is moving slower than the speed of light in vacuum. \emph{The electromagnetic field line curvature is non-singular in the vicinity of all charges}, even for point charges.

This means that in contrast to the conventional Maxwell-Lorentz formalism of classical electromagnetic theory, the covariant field lines formalism allows one to study the full electromagnetic field tensor via its field lines without any singularity issues. This holds independently of the structure of the charges, and continues to hold for point particles.


\subsection{Deriving an equation of motion from the curvature of the covariant field lines} \label{subsec:EOM}

This section shows the key result that the curvature of the field lines naturally yields the Lorentz force equation. The field line curvature in Eq. (\ref{eq:Curvature}) relates the known trajectory of a charge to the curvature of the field lines in the proximity of this trajectory. One may consider inverting this relationship in order to use the field line formalism to predict what will be the trajectory of the charge using this geometric approach. In Appendix \ref{sec:EOM} we consider this problem. The challenge is that there seem to be many routes to derive such an equation, and we considered only one approach as an application of our framework. The approach considered gave rise to the following equation of motion:
\begin{equation} \label{eq:EOM}
m a^\mu = q F^{\mu\nu}_{\text{ext}} u_\nu + m \tau_0 a^2 \left(\frac{\hat{k}^\mu}{\hat{R}} - u^\mu\right),
\end{equation}
the derivation of which assumes that $\hat{k}^\mu a_\mu=0$ (two other assumptions are discussed in Appendix $\ref{sec:EOM}$). Here $\hat{R} = -\hat{k}^\nu u_\nu$ and $\tau_0 =\frac{2 q ^2}{3m}$. For electrons, $q=e$ is the electron's charge, $\tau_0=6.24\times 10^{-24} \,\text{s}$, and the expansion parameter is one third of the classical electron radius. $\tau_0$ naturally appears when studying electromagnetic radiation \cite{bib:Jackson}.

Eq. (\ref{eq:EOM}) has many interesting properties. First, it conserves energy and is Lorentz invariant, as can be seen by contracting it with $u_\mu$. Second, by contracting it with $a_\mu$ we receive $ma^2 = q a_\mu F^{\mu\nu} _{\text{ext}} u_\nu$, which is a known result from the Lorentz force equation (note though that it may have an order $\tau_0 ^2$ corrections). In fact, when considering $\tau_0$ as a small parameter, we see that in physical systems with characteristic time frames that are considerably larger than $\tau_0$ (or alternatively the limit $\tau_0 \rightarrow 0$), Eq. (\ref{eq:EOM}) is nothing other than the Lorentz force equation $m a^\mu = q F^{\mu\nu}_{\text{ext}} u_\nu$ for a charge $q$ moving under an external field $F^{\mu\nu} _\text{ext}$. We remind that despite the fact that the initial field we considered was the total external and self field, the equation of motion derived is independent of the self field and thus its singularities, as was discussed in the previous section. We discuss the $\tau_0$-order terms in Eq. (\ref{eq:EOM}) next.


\section{Discussion} \label{sec:Summary}

In this paper we revisited the geometric origins of the electromagnetic theory as outlined by Faraday and Maxwell, and brought back the notion of field lines into the heart of the discussion.

We defined a notion of covariant electromagnetic field lines in Eq. (\ref{eq:CovariantFieldLines}), which naturally generalizes electric and magnetic field lines. We calculated the covariant field line curvature in Eq. (\ref{eq:VectorCurvature}) and derived an explicit closed-form formula for the curvature nearby an electric charge in motion (Eq. \ref{eq:Curvature}). Unlike most electromagnetic quantities, the electromagnetic field line curvature is not singular by the charge, and is even regular nearby a point charge.

Maxwell's theory predicts that when a charge accelerates, it loses energy due to electromagnetic radiation emission \cite{bib:Griffiths}. The rate at which energy is emitted by the charge in electromagnetic radiation is given by the relativistic Larmor formula, and more generally the rate at which energy-momentum is emitted is given by the Abraham-Heaviside formula $R^\mu = -m \tau_0 a^2 u^\mu$ \cite{bib:Jackson}. The loss of energy due to radiation is expected to produce a recoil force called the radiation reaction force, which one can identify as the last term in Eq. (\ref{eq:EOM}). The effect of radiation reaction is not present in the Lorentz force equation since the Lorentz force equation only includes the external electromagnetic force.

The precise nature of the radiation reaction force is still a matter of debate, which resisted a widely accepted solution for over a century. To date, several different models were suggested to correct the Lorentz force equation \cite{bib:Dirac1938nz, bib:Eliezer, bib:LandauLifshitz, bib:MoPapas, bib:Caldirola, bib:Yaghjian, bib:PrigogineHenin, bib:Nodvik, bib:Teitelboim, bib:PetzoldSorg, bib:MonizSharp, bib:LevineMonizSharp, bib:Rohrlich2, bib:GrallaHarteWald, bib:GalleyLeibovichRothstein}, but there does not seem to be a consensus about the validity of these models. While the majority of the work on radiation reaction is theoretical, the last decade marks a new opportunity to explore radiation reaction experimentally using high-intensity lasers. Both radiation reaction and self-force recently attracted much interest \cite{bib:PRX1,bib:PRX2,bib:PRX3,bib:NatCom18,bib:PRL18} alongside with other intriguing applications of high-intensity lasers \cite{bib:I1,bib:I2}. All the above can be quantitatively analyzed with the aid of our geometric framework and the resulting set of new equations of motion.

Eq. (\ref{eq:EOM}) greatly resembles the notorious Lorentz-Abraham-Dirac equation for radiation reaction $m a^\mu = qF^{\mu\nu} _{\text{ext}} u_\nu + m\tau_0 \left(\dot{a} ^\mu - a^2 u^\mu\right)$ except for one important difference \cite{bib:Dirac1938nz}. The notorious term $m\tau_0 \dot{a}^\mu$ in the Lorentz-Abraham-Dirac equation does not appear in Eq. (\ref{eq:EOM}). This notorious term results in many physical pathologies of the Lorentz-Abraham-Dirac equation, such as runaway solutions and pre-acceleration \cite{bib:Rohrlich}, both of which are not exhibited in Eq. (\ref{eq:EOM}).

When considering Eq. (\ref{eq:EOM}) as a potential equation of motion, two important questions rise. What is the interpretation $\hat{k}^\mu$ in the equation and most importantly how it may be used for calculation? We recall from section \ref{subsec:NoSingularities} that $\hat{k}$ is an arbitrary null-vector of the form  $\hat{k}^\mu=(1,\hat{k})$, where $\hat{k}$ is a unit-length three vector. In addition, the condition $\hat{W}=-\hat{k}^\mu a_\mu=0$ means that this null-vector $\hat{k}^\mu$ is perpendicular to the four-acceleration in spacetime. We note that this term is also proportional to a similar term in the quantum derivation of the Larmor formula (see Eq. (49) in \cite{bib:QuantumLarmor}). It may be that the interpretation of the $m \tau_0 a^2 \frac{\hat{k}^\mu}{\hat{R}}$ term in Eq. (\ref{eq:EOM}) as the momentum of the emitted photon can provide useful insight on recoil corrections to the equation of motion and relate to the equation of radiation reaction


Most importantly, the covariant field line curvature formulation allowed for a geometric derivation of the Lorentz force equation from first principles as well as its first order correction. The derivation of the Lorentz force helps to test the validity of the covariant field line approach, however the first order corrections are admittedly still speculative. A surprising property of the framework suggested here is that we started with the total electromagnetic field $F^{\mu\nu} = F^{\mu\nu} _\text{ext} + F^{\mu\nu} _\text{self}$, including the self-field, and derived the Lorentz force equation and its correction in Eq. (\ref{eq:EOM}), which is independent of the self-field singularity. In other words, the approach suggested here avoids the well-known challenges of self-force singularities.

Due to the assumptions in the derivation of Eq. (\ref{eq:EOM}), the correction to the Lorentz force is not expected to fully explain radiation reaction corrections, yet we hope that the approach will provide useful ideas for ways by which classical equations of motion could be modified and extended. The focus of this work is the study of covariant electromagnetic field lines as a new geometrical and physical object that may have various applications. Future applications may provide new insights into renormalization, e.g. by enabling alternative methods for second quantization in quantum electrodynamics. Other prospects include novel approaches to combining electromagnetic and gravitational field lines, as well as further ideas across different disciplines of physics such as the Yang-Mills theory \cite{bib:Hofmann}.

\section{Acknowledgements}
I.K. was supported by the Azrieli foundation as an Azrieli Fellow, and was partially supported by the Seventh Framework Programme of the European
Research Council (FP7-Marie Curie IOF) under grant no. 328853-MC-BSiCS and by the ERC starting grant NanoEP (no. 851780). E.C. was supported by the Israel Innovation Authority under project 70002, by the Pazy Foundations, by FQXi and by the Quantum Science and Technology Program of the Israeli Council of Higher Education.

\section{Appendix}
The Appendix includes additional details about the derivation of formulas regarding the geometry of electric and magnetic field lines in electrostatics and magnetostatics, as well as covariant electromagnetic field lines. The Appendix consists of five sections. In the first section, we define electric and magnetic field lines in three-dimensional space, calculate their Frenet-Serret vectors, use the vectors to compute the curvature of these field lines and prove that they imply the equivalence of curvature of field lines and acceleration of charges in electrostatics and magnetostatics. The second section discusses the uniqueness of the covariant field line definition employed in this work. The third section includes additional details on the derivation of covariant electromagnetic field lines and their curvature. The fourth section discusses the temporal and spatial components of the covariant field lines in the rest frame of an accelerating charge and their relationship to the three-acceleration of a charge. The fifth section derives an action principle for the dynamics of a charge using the field lines curvature. The sixth and last section demonstrates one way to derive a potential equation of motion from the field line formalism.

\subsection{Electric and Magnetic Field Lines} \label{sec:ElectrostaticFieldLines}

The field lines of a vector field $\vec{V}$ are the set of all curves in space which are tangent to the vector field at each point in space (in the mathematical literature, they are often called \emph{integral curves} \cite{bib:Lee}). Field lines can be found by solving \emph{the field line equation}
\begin{equation} \label{eq:FieldLines}
\frac{d \vec{x}}{d\lambda} = \vec{V}(\vec{x}(\lambda)),
\end{equation}
where $\vec{x}(\lambda)$ is a parametrization of each field line curve with a parameter $\lambda$. For example, the field $\vec{V}$ may be an electric field $\vec{V}=\vec{E}$, or a magnetic field $\vec{V}=\vec{B}$. With these two choices the field line Eq. (\ref{eq:FieldLines}) gives the electric field lines $\vec{x}_e$ and magnetic field lines $\vec{x}_m$ respectively,
\begin{equation} \label{eq:ElectricFieldLines}
\frac{d \vec{x}_e}{d\lambda} = \vec{E}(t, \vec{x}_e(\lambda)) \hspace{1cm} \frac{d \vec{x}_m}{d\lambda} = \vec{B}(t, \vec{x}_m(\lambda)).
\end{equation}
assuming a fixed time $t$. Since each field line is a curve in space, one may study its geometrical properties. In differential geometry in a three-dimensional space, a curve may be characterized by its three Frenet-Serret vectors: the tangent vector $\vec{T}$, the normal (curvature) vector $\vec{N}$ and its binormal vector $\vec{B}$ \cite{bib:Spivak}. The Frenet-Serret vectors of a general vector field $\vec{V}$ can be expressed explicitly as
\begin{eqnarray} \label{eqs:FrenetSerret}
\vec{T} &=& \frac{\vec{V}}{|\vec{V}|}, \\ \notag
\vec{N} &=& \frac{|\vec{V}|^2 (\vec{V}\cdot\nabla)\vec{V} - \left(\vec{V}\cdot(\vec{V}\cdot\nabla)\vec{V}\right)\vec{V}}{|\vec{V}| \cdot |\vec{V} \times (\vec{V}\cdot \nabla)\vec{V})|}, \\ \notag
\vec{B} &=& \frac{\vec{V}\times (\vec{V}\cdot \nabla) \vec{V}}{|\vec{V}\times (\vec{V}\cdot \nabla) \vec{V}|}.
\end{eqnarray}
The Frenet-Serret vectors $\vec{T}$, $\vec{N}$ and $\vec{B}$ are an orthonormal set, as can be verified directly using Eqs. (\ref{eqs:FrenetSerret}). One may substitute $\vec{V}=\vec{E}$ or $\vec{V}=\vec{B}$ to obtain a general expression for the Frenet-Serret vectors of the electric and magnetic field respectively, which will not be provided here.

In this paper, we are interested in the curvature of the electric and magnetic fields. For a general vector field $\vec{V}$, the curvature of each field line is given by the formula
\begin{equation} \label{eq:FSCurvature}
\kappa = \frac{\vec{T} ' \cdot \vec{N}}{|\vec{V}|},
\end{equation}
where prime denotes a derivative with respect to the parameter $\lambda$. Substituting Eqs. (\ref{eqs:FrenetSerret}) into Eq. (\ref{eq:FSCurvature}) gives a formula for the curvature of the field lines at each point in space
\begin{equation} \label{eq:Curvature3D}
\kappa = \frac{|\vec{V}\times(\vec{V}\cdot\nabla) \vec{V}|}{|\vec{V}|^3}.
\end{equation}

Consider now a charge $q$ traveling along the path $\vec{x}_0 (t)$ under the influence of an external electric field $\vec{E}_{\text{ext}}$. The charge produces an additional \emph{self} field $\vec{E}_{\text{self}}$, and the total field is the superposition $\vec{E}=\vec{E}_{\text{self}} + \vec{E}_{\text{ext}}$. From Eq. (\ref{eq:Curvature3D}) the electric field lines curvature is
\begin{equation} \label{eq:ElectricCurvature}
\kappa(t, \vec{x}) = \frac{|\vec{E} \times (\vec{E} \cdot \nabla) \vec{E}|}{|\vec{E}|^3}.
\end{equation}
One may expand Eq. (\ref{eq:ElectricCurvature}) to a power series, and in the leading order nearby the charge it gives \cite{bib:Hadad1}
\begin{equation} \label{eq:ElectrostaticCurvature}
\kappa(t, \vec{x}) = \frac{3}{q} |\vec{E} _{\text{ext}}(t, \vec{x}) \times (\vec{x}_0 (t) - \vec{x})|.
\end{equation}

If we choose a coordinate system in which the charge is instantaneously at rest (the charge's instantaneous rest frame), the infinitesimal displacement of the charge is $\vec{x}_0(t+\Delta t) - \vec{x}_0(t) = \frac{1}{2} \vec{a}_0 (t) (\Delta t)^2$, where $\vec{a}_0(t)$ is the charge acceleration. The electric field lines curvature Eq. (\ref{eq:ElectrostaticCurvature}) can be evaluated along the direction the charge will accelerate to, yielding
\begin{equation}
\kappa(t, \vec{x}_0 (t+ \Delta t)) = \frac{3 (\Delta t)^2}{2q} |\vec{E} _{\text{ext}}(t, \vec{x}_0 (t + \Delta t)) \times \vec{a}_0 (t)|.
\end{equation}
One can readily notice that the curvature is zero $\kappa(t, \vec{x} (t+ \Delta t)) = 0$ if and only if the acceleration is parallel to the external field. Therefore \emph{an electric charge always accelerates in the direction of a straight electric field line}. A similar computation can be used to show that the same result holds for a charge in the presence of a magnetic field.



\subsection{Uniqueness of the Covariant Electromagnetic Field Lines} \label{sec:Uniqueness}

Observing the form of the electric and magnetic field line Eqs. (\ref{eq:3DFieldLines}), it is natural to expect a covariant electromagnetic field line equation of the form
\begin{equation} \label{eq:ElectromagneticFieldLine}
\frac{d x ^\mu}{ds} = G^\mu(x^\nu (s)),
\end{equation}
in Minkowski spacetime where $x^\mu (\lambda)$ is the electromagnetic field line parameterized by $\lambda$ and the non-homogeneous term $G^\mu$ provides the dependency on the electromagnetic field.

We look for $G^\mu$ that will be linear in the field tensor $F^{\mu\nu}$ such that its spatial part would reproduce Eqs. (\ref{eq:3DFieldLines}) in the limit of $\vec{B} \rightarrow 0$ or $\vec{E}\rightarrow 0$ respectively. Because we expect $G^\mu$ to be linear in $F^{\mu\nu}$ we have $G^\mu \sim F^{\mu\nu} \xi_\nu$ where $\xi_\nu$ is a four-vector. We want $\xi_\nu$ to be such that in the magnetostatic limit $B\rightarrow 0$ and $u^\mu\rightarrow 0$ we get the electric field line, i.e. the first of Eqs. (\ref{eq:ElectricFieldLines}). Thus $\xi_\nu$ must depend on the four-velocity of the charge linearly. It is easy to see that if we set
\begin{equation} \label{eq:ElectricField4D}
G^\mu (x^\nu) = F^{\mu\gamma} (x^\nu) u_\gamma (\tau_\text{ret}),
\end{equation}
namely, $G^\mu$ is the electromagnetic field at a spacetime event $x^\nu$ as measured by a charge traveling with the world line $z^\mu(\tau)$, where $u^\mu=\frac{dz^\mu}{d\tau}$ is the four-velocity of the charge, and $\tau_\text{ret}$ is the retarded time from the charge to the spacetime event $x^\nu$, then the spatial components of Eq. (\ref{eq:ElectromagneticFieldLine}) with the definition in Eq. (\ref{eq:ElectricField4D}) reduce to the electric and magnetic field line Eqs. (\ref{eq:3DFieldLines}). $\tau_\text{ret}$ must be used in Eq. (\ref{eq:ElectricField4D}) in order to maintain Lorentz covariance.

We will therefore refer to Eq. (\ref{eq:ElectromagneticFieldLine}) with Eq. (\ref{eq:ElectricField4D}) as \emph{the covariant electromagnetic field line equation}:
\begin{equation}
\frac{d x ^\mu}{ds} = F^{\mu\gamma} (x^\nu) u_\gamma (\tau_\text{ret}).
\end{equation}


\subsection{On the Derivation of the Covariant Electromagnetic Field Line Curvature} \label{sec:CovariantFieldLinesDerivation}


The curvature of covariant electromagnetic field lines is given by Eq. (\ref{eq:VectorCurvature}).
A general electromagnetic field can be written as the superposition $F^{\mu\nu} = F^{\mu\nu} _\text{ext} + F^{\mu\nu} _\text{self}$ of the external field $F^{\mu\nu} _\text{ext}$ and the self field $F^{\mu\nu} _\text{self}$ produced by the charge of interest on its forward light-cone. We substitute the self electric field $F^{\mu\nu} _\text{self}$ into Eq. (\ref{eq:ElectricField4D}) to obtain
\begin{equation} \label{eq:SelfElectricField}
E_{\text{self}} ^\mu = F^{\mu\gamma} _\text{self} (x^\nu) u_\gamma (\tau_\text{ret}) = -\frac{q}{R} U^\mu + \frac{q}{R^3}(1-W) k^\mu,
\end{equation}
with squared Minkowski norm
\begin{equation} \label{eq:SelfElectricFieldNorm}
(E_{\text{self}}) ^2 = \frac{q^2}{R^4} (1-W^2+a^2 R^2).
\end{equation}

Substituting $F^{\mu\nu} = F^{\mu\nu} _\text{ext} + F^{\mu\nu} _\text{self}$ into Eq. (\ref{eq:VectorCurvature}) shows that one needs to differentiate the self electric field Eq. (\ref{eq:SelfElectricField}) with respect to $x^\nu$. We proceed by taking the derivative $\frac{\partial}{\partial x^\nu}$ of the light cone Eq. (\ref{eq:kNull}),
\begin{equation} \label{eq:kNull}
k^\mu k_\mu = 0,
\end{equation}
which yields
\begin{equation}
\frac{\partial\tau_{\text{ret}}}{\partial x^\nu} = -\frac{k_\nu}{R}.
\end{equation}
This means that the derivative of $k^\mu$ can be written explicitly as
\begin{equation} \label{eq:dk}
\frac{\partial k^\mu}{\partial x^\nu} = \delta^\mu _\nu + \frac{u^\mu k_\nu}{R}.
\end{equation}
The derivative of the retarded distance $R$ is
\begin{equation} \label{eq:dR}
\frac{\partial R}{\partial x^\nu} = -u_\nu + B k_\nu,
\end{equation}
and the derivative of $W$ is
\begin{equation} \label{eq:dW}
\frac{\partial W}{\partial x^\nu} = -a_\nu - \frac{A k_\nu}{R}
\end{equation}
where $A$ is the retarded acceleration which is defined in terms of the jerk $\dot{a}^\mu$ (the third time derivative of the four-position) as
\begin{equation} \label{eq:A}
A = -k \cdot \dot{a}.
\end{equation}
The derivatives of the four-velocity and four-acceleration are
\begin{equation} \label{eq:du}
\frac{\partial u^\mu}{\partial x^\nu} = -\frac{a^\mu k_\nu}{R},
\end{equation}
and
\begin{equation} \label{eq:da}
\frac{\partial a^\mu}{\partial x^\nu} = -\frac{\dot{a}^\mu k_\nu}{R}
\end{equation}
respectively.

Using the identities (\ref{eq:dk}--\ref{eq:da}) in Eq. (\ref{eq:VectorCurvature}) gives the covariant electromagnetic field lines curvature, as presented in Eq. (\ref{eq:Curvature}) in the main text. The easy-to-verify identities $\partial_\mu U^\mu = U^\mu \partial_\mu R = 0$ and $U^\mu k_\mu = -1$, are helpful in the calculation.

\subsection{The Components of the Covariant Field Lines in the Rest Frame of an Accelerating Charge} \label{sec:FieldLinesComponents}

For completeness of the current manuscript, we follow here our previous analysis presented in \cite{bib:Hadad2}. Consider the motion of a charge in its instantaneous rest frame. The four-velocity is $u^\mu=(1,0,0,0)$ and the four-acceleration is $a^\mu=(0,\vec{a})$, where $\vec{a}$ is the spatial acceleration of the charge. The electromagnetic field lines four-curvature in Eq. (\ref{eq:VectorCurvatureFreeCharge}) from the main text can be written as
\begin{equation}
\kappa ^\mu = |\vec{a}|(-\cos \theta,-\hat{a} + \hat{k} \cos \theta).
\end{equation}
The scalar curvature is equal to the squared norm of the curvature four-vector,
\begin{equation}
\kappa^2 = -\vec{a}^2 \cos(2\theta),
\end{equation}
where $\theta$ is the spatial angle between the acceleration $\vec{a}$ and the retarded position of the charge $\vec{k}=\vec{x}-\vec{z}(\tau_\text{ret})$(when both are represented as 3-vectors). When we separate the four-curvature $\kappa^\mu$ to its temporal ($0$-component) and spatial parts (components $1,2,3$) by writing $\kappa^\mu = (\kappa^0, \vec{\kappa})$, we can represent the \emph{scalar spatial curvature} as 
\begin{equation} \label{eq:SpatialCurvature}
|\vec{\kappa}| = |\vec{a}| |\sin \theta|.
\end{equation}
Hence, the field lines of an accelerating charge are locally straight in space only in the direction of the acceleration $\theta=0$. In such cases the temporal component of the field line curvature is equal to the magnitude of the acceleration (up to a minus sign)
\begin{equation}
\kappa^0 = -|\vec{a}|.
\end{equation}
Therefore, in the vicinity of a freely accelerating charge, its kinematic properties are fully determined by the field line curvature. This extends the result derived in \cite{bib:Hadad1}.

\subsection{Action Principle for Electrostatic Field Lines} \label{sec:ActionPrinciple}

The formalism provided in this work gives a novel geometrical framework for electrodynamics which touches upon the issues of radiation reaction and self-force. Since both of these problems are related to the joint conservation of energy fields and massive particles, it is interesting to explore the conservation laws implied by our geometrical approach. We can study the conservation of energy by attempting to define an action principle for the particle's motion. It is supposed to be consistent since our approach does not suffer from divergence of the field at the position of the particle. However, our geometric description addresses at the moment the dynamics of particles, but not the dynamics of fields, hence one cannot expect to see, at least in this stage, conservation of the total energy. The goal of this section is to take the first step in this direction and show that in the case of electrostatic field lines the answer is affirmative and one can derive indeed an action principle based on the field lines curvature.

To see this, note that in the vicinity of the charge the non-relativistic curvature vector is
\begin{equation}
\vec{k} = \frac{3}{q} \vec{E}_\text{ext} (\vec{x}) \times (\vec{x}-\vec{x}_0 (t)).
\end{equation}
If there is nearly no induction (or no magnetic fields) $\nabla \times \vec{E} \approx \vec{0}$ and
\begin{equation}
\nabla \times \vec{k} = \frac{6}{q} \vec{E} _\text{ext} (\vec{x}).
\end{equation}
This means that the scalar electric potential can be expressed using the curvature as
\begin{equation} \label{eq:ElectricPotential}
\phi(\vec{x}) = -\frac{q}{6} \int ^{\vec{x}} (\nabla \times \vec{k})\cdot d\vec{l}.
\end{equation}
We therefore obtain an action principle for the dynamics of the charge with the Lagrangian
\begin{equation} \label{eq:NonRelativisticLagrangian}
L = \frac{1}{2} mv^2 - \phi,
\end{equation}
where $\phi$ is defined in Eq. (\ref{eq:ElectricPotential}) using the field lines curvature alone. Since the curl operator $\nabla \times$ measures the circulation density of a field, Eq. (\ref{eq:NonRelativisticLagrangian}) can be given a simple interpretation. The action principle teaches us that \emph{a charge would travel along the path of least electric curvature circulation}. This result is important, showing that the entire dynamics in this case can be derived by the field lines curvature alone. It should be possible to employ such analysis in the general electromagnetic case and to include within it the dynamics of fields as well.

\subsection{Deriving the Equation of Motion from the Field Line Formalism} \label{sec:EOM}

This work analyzed the field line curvature $\kappa^\mu$ as a vector field that is defined throughout spacetime. We also considered a given trajectory of a specific charge, to derive the field line curvature in the proximity of the trajectory in Eq. (\ref{eq:Curvature}). This equation gives an explicit way to calculate the curvature of the field lines in a little tube that surrounds the known trajectory of the charge, and uses the external field $F^{\mu\nu} _{\text{ext}}$ and properties that depend on the trajectory of the charge, such as its four-velocity, four-acceleration, and jerk. In this sense, the field line curvature Eq. (\ref{eq:Curvature}) gives a way to transform a known world line and electromagnetic fields to get the field lines curvature. If the trajectory of the charge is known up to a certain point on its world line, the field line curvature can still be calculated nearby the known section of the worldline.

We now want to use this information in order to invert it. Namely, we hope to use this known relationship, mapping the external field and the trajectory to the field line curvature, in order to predict how the trajectory of a charge will continue beyond a known point. In other words, we wish to derive the equation of motion using this process. There seem to be different ways to mathematically derive a potential equation of motion from the field line formalism. The exact derivation and end result would very much depend on the assumptions that are considered legitimate and that one wishes to explore. In this sense, this section considers a few speculative assumptions that give rise to an equation that very much resembles known equations. This is by no means the unique way to do so, and is merely used as a way to demonstrate how one may take advantage of the field line formalism.

Consider the curvature of field lines in Eq. (\ref{eq:Curvature}). We wish to use it in order to derive an equation of motion for the charge moving through the field lines. We expect an equation of motion that recreates the Lorentz force equation in the right limit. The only term in Eq. (\ref{eq:Curvature}) that is proportional to the Lorentz force equation is the last term, and we require it to be equal to the Lorentz force, namely $-\frac{3 \varepsilon \hat{R}}{q} F^{\mu\nu}_{\text{ext}} u_\nu = \frac{q}{m} F^{\mu\nu}_{\text{ext}} u_\nu$ along the trajectory of the charge. This condition requires the expansion parameter to satisfy $\hat{R}\varepsilon = -\frac{1}{2}\tau_0$, with $\tau_0 =\frac{2 q ^2}{3m}$. For electrons, we substitute $q=e$ as the electron's charge, $\tau_0=6.24\times 10^{-24} \,\text{s}$, and the expansion parameter is one third of the classical electron radius. $\tau_0$ naturally appears when studying electromagnetic radiation \cite{bib:Jackson}. Under this assumption, we may rewrite Eq. (\ref{eq:Curvature}) as
\begin{equation} \label{eq:CurvatureTau0}
\begin{array}{l}
a^\mu - \frac{q}{m} F^{\mu\nu}_{\text{ext}} u_\nu + \kappa^\mu = -\frac{\hat{W}}{\hat{R}^2} \hat{P}^\mu + \frac{\hat{W}}{\hat{R}} u^\mu +\\
\hspace{.8cm} -\frac{\tau_0}{2\hat{R}} \left[\left(\frac{2\hat{W}^2}{\hat{R}^2} - 2a^2-\frac{3}{q\hat{R}} \hat{k}_\nu F_{\text{ext} }^{\nu\gamma} u_\gamma + \frac{\hat{A}}{\hat{R}}\right)\hat{P}^\mu-\left(\frac{\hat{W}^2}{\hat{R}} u^\mu + \hat{W} a^\mu + \hat{R} \dot{a}^\mu \right)\right]
\end{array}
\end{equation}
In time frames much larger than $\tau_0$, one may consider $\tau_0$ as a parameter for perturbation theory, and taking the limit $\tau_0 \rightarrow 0$ in Eq. (\ref{eq:CurvatureTau0}) restores the familiar Lorentz force equation $m a^\mu = e F^{\mu\nu} _{\text{ext}} u_\nu$.

By contracting this equation with $u_\mu$ (recall that $F^{\mu\nu}$ is anti-symmetric, $u_\mu a^\mu=0$, $\hat{P}^\mu u_\mu=0$, and $u^\mu u_\mu = -1$), we see it must satisfy the condition
\begin{equation}
    \kappa^\mu u_\mu = -\frac{\hat{W}}{\hat{R}}+\frac{\tau_0}{2\hat{R}}\left(-\frac{\hat{W}^2}{\hat{R}}+\hat{R}\dot{a}^\mu u_\mu\right).
\end{equation}
If we stick to the condition that was studied in section \ref{subsec:CurvatureIsAcceleration}, $\hat{W}=0$, and we need to require that
\begin{equation}
    \kappa^\mu u_\mu = \frac{1}{2} \tau_0 \dot{a}^\mu u_\mu.
\end{equation}
It can be shown that this is equivalent to stating that the projection of the curvature on the four-momentum to be equal to half the rate of energy emission by radiation, namely $\kappa^\mu p_\mu = -\frac{1}{2} m \tau_0 a^2$. This condition holds if we take $\kappa^\mu=\frac{1}{2}\tau_0 \dot{a}^\mu$. Under these two assumptions on $\kappa$ and $\hat{W}$, Eq. (\ref{eq:CurvatureTau0}) gives
\begin{equation}
m a^\mu = q F^{\mu\nu}_{\text{ext}} u_\nu + m \tau_0 \left[\left(\frac{a^2}{\hat{R}} + \frac{3}{2 q\hat{R}^2} \hat{k}_\nu F_{\text{ext} }^{\nu\gamma} u_\gamma - \frac{\hat{A}}{2\hat{R}^2}\right)\hat{P}^\mu\right]
\end{equation}

Contracting this equation with $k_\mu$, we note that that because $\hat{k}^\mu a_\mu = -\hat{W}=0$, both $\hat{W}$ and its derivative $\hat{A}$ are order $\tau_0$, and thus contribute a term of order $(\tau_0) ^2$ which we omit. This means that \emph{given the assumptions provided}, we may consider the limit $\tau\rightarrow \tau_{\text{ret}}$ and may conclude that up to order $\tau_0$ the trajectory of the charge must satisfy
\begin{equation}
m a^\mu = q F^{\mu\nu}_{\text{ext}} u_\nu + m \tau_0 a^2 \left(\frac{\hat{k}^\mu}{\hat{R}} - u^\mu\right)
\end{equation}
One may consider this as an equation of motion for the trajectory of the charge. The exact interpretation of this equation and its limitations are discussed in the discussion section.

%
%


\end{document}